# Quantum sensing-enabled deuterium NMR spectroscopy with nanoscale sensitivity at low magnetic fields

Dileep Singh[1,2], Riley W. Hooper[1,2], Christoph Findler[3], Utsab Banerjee[1,2], Dominik B. Bucher[1,2*]

[1]Technical University of Munich, TUM School of Natural Sciences, Chemistry Department, Lichtenbergstraße 4, 85748 Garching bei München, Germany

[2]Munich Center for Quantum Science and Technology (MCQST), Schellingstr. 4, 80799, München, Germany

[3]Diatope GmbH, DLR-Building 3, Wilhelm-Runge-Straße 10, 89081 Ulm, Germany

*Corresponding author: dominik.bucher@tum.de

**ABSTRACT**

Nuclear magnetic resonance (NMR) spectroscopy provides unparalleled access to molecular structure and dynamics but is traditionally limited by weak signal strength, requiring large sample volumes and high magnetic fields. Here, we demonstrate nanoscale deuterium ($^2$H) NMR spectroscopy using nitrogen-vacancy (NV) centers in diamond, reproducing the characteristic quadrupolar powder line shapes that are present in the conventional bulk NMR spectra. By detecting statistical spin fluctuations from nanometer-scale detection volumes, our approach delivers a sensitivity enhancement of six to eight orders of magnitude over inductive detection while operating at magnetic fields two orders of magnitude lower than those used in conventional NMR. Temperature-dependent measurements of a deuterated polymer and molecular solid reveal distinct motional averaging and phase transitions with nanoscale sensitivity. Powder-like NV-detected $^2$H NMR establishes a powerful tool for probing molecular dynamics on the nanoscale and, in the ultimate limit, at the single-molecule level - capabilities beyond those of most existing spectroscopic techniques.

**INTRODUCTION**

NMR spectroscopy is one of the most powerful analytical techniques for probing molecular structure and dynamics. However, NMR signals are intrinsically weak because nuclear spin polarization is extremely small at thermal equilibrium, making high magnetic fields necessary to generate detectable signal with conventional (bulk) NMR (*1, 2*). The NMR signal can be enhanced by orders of magnitude through dynamic nuclear polarization (DNP), which transfers the much larger electron spin polarization to nuclear spins and thereby provides substantial sensitivity gains (*3–6*). An alternative way to measure NMR signals with improved sensitivity is to use a novel type of sensor – the electronic spin of nitrogen-vacancy (NV) centers in diamond (*7–16*). These optically addressable solid state spin defects function as nanoscale magnetometers (*17–20*). When positioned only a few nanometers below the diamond surface, they are sensitive to the statistical polarization arising from nanoscale volumes of spin-active nuclei, thereby circumventing the need for inductive detection of large ensembles of Boltzmann polarized spins (*7, 8, 21–23*). Spin noise detection offers several advantages that are particularly relevant for solid state samples: it eliminates the need to excite nuclear spins with RF fields, requires no longitudinal spin ($T_1$) recovery between measurements and provides a signal strength that is independent of the applied magnetic field (*21*, *24–26*). Together with nanoscale detection volumes, this capability has enabled the detection of spins at interfaces (*27–29*), in single molecules (*30*), and down to the level of individual nuclear spins (*31*, *32*).

NV-based quantum sensing protocols are most efficient at low magnetic fields (< 1 T) and low detection frequencies (up to a few tens of megahertz) (*33*, *34*). Moreover, nanoscale NV NMR is intrinsically constrained by limited spectral resolution, particularly in solids, where strong dipolar couplings lead to substantial line broadening and where magic-angle spinning is difficult to implement (*9*, *35*).

Nuclei with spin $I > ½$ can transform these apparent limitations into an advantage, as their characteristic quadrupolar splittings generate distinct, information-rich, and intrinsically broad spectral features, even at low magnetic fields. These intrinsically broad resonances naturally match the spectral resolution currently accessible with NV-based techniques. This has been impressively demonstrated by using NV centers to probe quadrupolar nuclei such as $^{10/11}$B and $^{14}$N in single hexagonal boron nitride flakes, a two-dimensional material (*36*, *37*).

Deuterium ($^2$H) presents a prototypical quadrupolar probe, exhibiting Pake patterns with characteristic line shapes on the order of several tens of kilohertz that are highly sensitive to local structural dynamics (*38*). These powder patterns provide unique spectroscopic access to molecular geometry, orientational order parameters, and motional timescales spanning a broad dynamical range, from fast intramolecular rotations to slow conformational motions and chemical exchange. While Lovchinsky *et al.* (*30*) demonstrated NV-based detection of $^2$H signals in single proteins, no quadrupolar splitting was resolved in their measurements. To date, nanoscale NV-based NMR has not enabled the observation of distinct powder-like quadrupolar NMR line shapes.

Here we use an NV ensemble which allows us to detect $^2$H spectra that closely match conventional bulk solid state $^2$H NMR – a long-standing goal in nanoscale NV-NMR. The ultrasmall detection volumes translate into an effective sensitivity enhancement exceeding six to eight orders of magnitude over standard inductive detection. Remarkably, this performance is achieved at magnetic fields two orders of magnitude lower than those typically needed for high resolution quadrupolar spectroscopy in bulk NMR, where ultrahigh fields are conventionally used to boost sensitivity and spectral resolution. By comparing bulk and nanoscale NV-NMR spectra recorded at different temperatures, we extract information on molecular dynamics and phase transitions in both a polymer and a molecular system. To the best of our knowledge, this represents the first NV-based NMR approach to deliver chemical information in the form of a quadrupolar Pake pattern comparable to conventional bulk measurements, while simultaneously eliminating the need for high magnetic fields and providing dramatically enhanced sensitivity. Our results open the door to probing molecular dynamics on the nanoscale, ranging from interfaces down to single molecules.

**RESULTS**
**Nanoscale $^2$H NV-NMR**

Nanoscale detection of $^2$H nuclear spins is achieved using a shallow ensemble of NV centers, implanted a few nanometers (~8 nm) below the surface of an electronic-grade diamond chip (Fig. 1A). Achieving sufficient sensitivity for detecting $^2$H is challenging due to its moderate gyromagnetic ratio, which is approximately 6.5 times smaller than that of $^1$H, resulting in ~40x reduction in signal strength. To overcome this limitation, we developed an improved NV-diamond platform optimized for enhanced spin sensitivity (detailed in MATERIALS AND METHODS). A layer of deuterated material, such as PMMA-d$_8$, was deposited directly onto the diamond surface with gentle thermal treatment to ensure uniform diamond coverage (see MATERIALS AND METHODS for details). Optical initialization and readout of the NV spin state is performed using green laser excitation, while coherent spin control is achieved via a microwave loop positioned on the diamond surface. Due to the tetrahedral symmetry of the diamond lattice, NV centers occur in four crystallographic orientations, of which the subset aligned with an external magnetic field (B$_0$) constitutes the effective sensing ensemble.

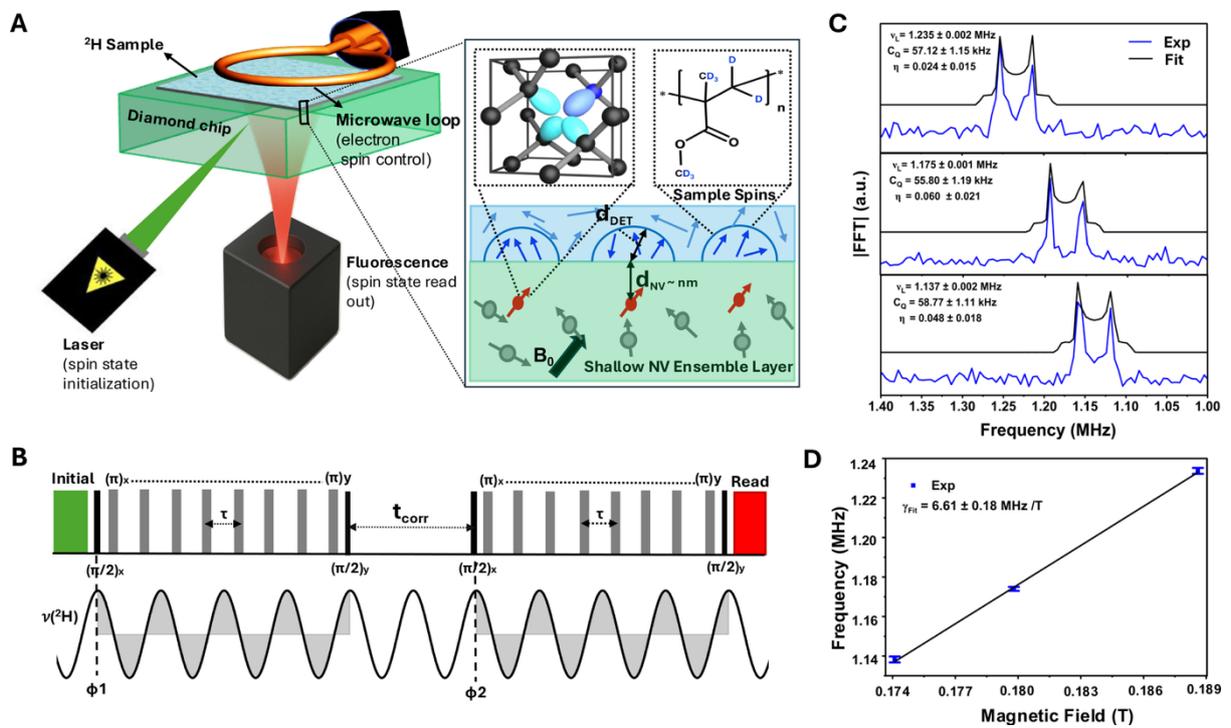

**Figure 1. Nanoscale $^2$H NMR using NV centers. (A)** A deuterated sample is deposited on a diamond chip hosting a shallow ensemble of NV centers. Optical excitation initializes and reads out the NV electronic spins, while microwave fields provide coherent spin control. The enlarged insets illustrate the nanoscale sensing geometry: $^2$H spins in the deuterated polymer sample couple to near-surface NV centers via magnetic dipolar interactions, with the detection volume defined by the NV–sample separation d$_{NV}$ (a few nanometers). An external magnetic field B$_0$ along the NV axis sets the $^2$H Larmor frequency. The insets highlight the NV defect in diamond and the deuterated PMMA-d$_8$. **(B)** Pulse sequence used for NV-based $^2$H correlation spectroscopy. The NV electron spin is initialized optically, followed by a dynamical decoupling block composed of equally spaced π pulses with inter pulse delay τ, which filters magnetic noise at the $^2$H Larmor frequency. Two identical sensing blocks are separated by a variable time t$_{corr}$, correlating the nuclear spin fluctuations with the NV interaction window. The relative phases ϕ1 and ϕ2 of the nuclear precession during the two blocks determine the accumulated NV phase and hence the measured signal, which is read out optically at the end of the sequence. **(C)** Experimental $^2$H spectra with fitted spectra at different magnetic fields, yielding coupling constants C$_Q$ ≈ 55–59 kHz and asymmetry parameter η ≈ 0.024–0.060. **(D)** Magnetic field dependence of the detected $^2$H resonance frequency, showing a linear fit with a slope consistent with the gyromagnetic ratio of deuterium.

Near surface NV centers sense statistical magnetic field fluctuations from $^2$H nuclear spin dynamics within approximately a hemispherical detection volume ($d_{DET}$) defined by the NV depth $d_{NV}$ (*7*, *8*). Spectroscopic detection of the $^2$H signal is performed using an NV-based correlation spectroscopy protocol (*39–41*) built from dynamical decoupling pulse sequences (Fig. 1B) (*42*). The NV spin is interrogated using two identical XY8-*N* pulse blocks separated by a variable correlation time $t_{corr}$, with each block tuned to the $^2$H Larmor frequency. This sequence selectively filters magnetic noise at the Larmor frequency while suppressing off-resonant fluctuations. During the free evolution period between the two pulse blocks, the $^2$H nuclear spins precess and generate oscillatory magnetic fields picked up by the NV electronic spin. By sweeping $t_{corr}$, the protocol correlates these time-dependent magnetic field fluctuations at two distinct time points, producing oscillations in the NV photoluminescence (PL) signal. These oscillations directly encode the $^2$H Larmor precession, resembling the free induction decay in conventional inductive NMR, and are Fourier-transformed to yield the nanoscale $^2$H NMR spectrum. For the detection of statistical spin polarization, no nuclear spin excitation is required. Figure 1C shows experimentally measured $^2$H NV-NMR spectra recorded under three different magnetic field strengths around 180 mT. The spectra exhibit well-defined quadrupolar splittings characteristic of $^2$H nuclei in the PMMA-$d_8$ polymer, and the resonance positions shift systematically with field, corresponding to distinct $^2$H Larmor frequencies near 1.2 MHz. To unambiguously identify the nuclear species, the extracted resonance frequencies are plotted as a function of the applied magnetic field strength. The data exhibit a linear Zeeman dependence, yielding an experimental gyromagnetic ratio of $\gamma = 6.61\pm0.18$ MHz T$^{-1}$, which agrees well with the known value for deuterium ($\gamma = 6.552$ MHz T$^{-1}$), confirming that the detected signal originates from $^2$H nuclear spins (Fig. 1D).

**Benchmarking $^2$H nanoscale NV-NMR against conventional bulk NMR**
Next, we benchmark our NV-NMR measurements of two deuterated materials (PMMA-$d_8$, phenanthrene-$d_{10}$) with conventional bulk measurements acquired at 11.75 T (Fig. 2). Despite operating at a magnetic field ($B_0$) of only 188 mT, corresponding to a $^2$H Larmor frequency of ~1.2 MHz, the NV-NMR spectra exhibit well-resolved quadrupolar powder patterns with sufficient signal-to-noise to reliably extract the quadrupolar coupling constant $C_Q$ and asymmetry parameter η. The ability to reproduce bulk-like quadrupolar line shapes at very low magnetic fields demonstrates the exceptionally high detection efficiency and sensitivity advantage of NV-NMR over conventional NMR.

The bulk $^2$H spectrum for PMMA-$d_8$ (Fig. 2B) features two overlapping resonances originating from –CD$_3$ ($C_Q \approx 52$ kHz) and –CD$_2$ ($C_Q \approx 165$ kHz) groups, agreeing with earlier reports from $^2$H NMR literature (*43*). The NV-NMR spectrum for PMMA-$d_8$ (Fig. 2A) closely resembles the corresponding bulk $^2$H NMR spectrum (Fig. 2B), with a comparable (although slightly higher) quadrupolar coupling value. The NV-based measurement achieves a $^2$H spin sensitivity of ~8×10$^{13}$ spins √Hz$^{-1}$, compared to ~3×10$^{20}$ spins √Hz$^{-1}$ for bulk inductive NMR, corresponding to an improvement in spin sensitivity of more than six orders of magnitude.

Next, we probe a molecular solid, phenanthrene-$d_{10}$, for which the NV-NMR spectrum shows a reduced effective quadrupolar coupling constant (Fig. 2C) compared to the bulk NMR measurement (Fig. 2D). In the bulk case, a broad quadrupolar Pake pattern is observed with poor SNR and a large frequency shift. An additional, lower-intensity $^2$H signal ($C_Q < 400$ Hz) also appears around $\Delta\nu \approx 0.5$ kHz for phenanthrene-$d_{10}$ when a Hahn-echo pulse sequence is applied, whose identity could not be confirmed (Fig S1). The large $^2$H frequency shift, extreme broadening, and poor SNR for the spectrum of the main signal in Fig. 2D is attributed to the presence of a persistent radical species impacting the phenanthrene-$d_{10}$ NMR spectrum, which has been confirmed via EPR spectroscopy (Fig. S2) (*44–47*). NV-NMR measurements of

phenanthrene-d$_{10}$ further exhibit substantially higher sensitivity than conventional bulk NMR, corresponding to an improvement of approximately eight orders of magnitude, exceeding the sensitivity enhancement observed for PMMA-d$_8$.

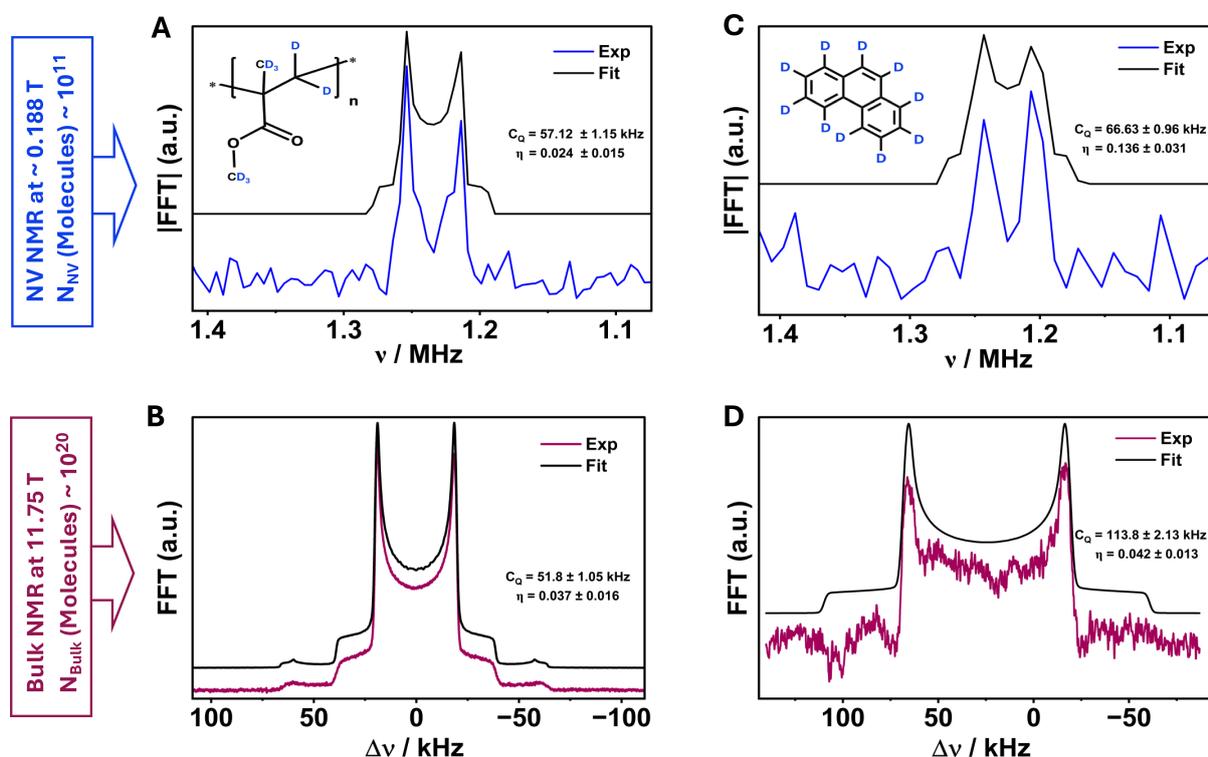

**Figure 2. Benchmarking of NV-detected and bulk $^2$H NMR spectra.** $^2$H NV-NMR spectra of a PMMA-d$_8$ thin film (**A**) are compared with the corresponding bulk spectrum (**B**), showing similar quadrupolar powder patterns and comparable quadrupolar coupling constants C$_Q$ and asymmetry parameters η, but with 7 orders of magnitude higher sensitivity at much lower magnetic field. For phenanthrene-d$_{10}$, the NV-NMR spectrum (**C**) displays a clear quadrupolar splitting, with the bulk spectrum (**D**) presenting an even larger quadrupolar-broadened line shape. Experimental spectra (blue, purple) are overlaid with simulated spectra (black) fit using ssNake to extract C$_Q$ and η.

To rationalize the observed quadrupolar couplings and their differences between polymer and molecular systems, we compare the experimental NV-NMR and bulk NMR results with density-functional theory (DFT) calculations for the static quadrupolar NMR parameters. DFT calculations were performed on molecular models of all three compounds using ORCA 6.1 (see MATERIALS AND METHODS and Fig. S4-S5, Tables S3-S4 in SI) to extract the static quadrupolar coupling constants and asymmetry parameters. The large C$_Q$ values (≈ 187–197 kHz) for all compounds predicted by DFT reflect rigid-lattice electric field gradients (EFGs) in the absence of motion, whereas the experimentally observed couplings are reduced by the presence of molecular dynamics. Table 1 compares quadrupolar parameters extracted from $^2$H NV-NMR, bulk NMR, and DFT calculations, revealing a clear distinction between static structure EFGs and system-dependent dynamic averaging of the quadrupolar interaction.

**Table 1.** Fit EFG parameters for deuterated samples as measured by NV-NMR, conventional bulk NMR, and calculated using ORCA 6.1.0 (DFT). Reported DFT values are the average of the quadrupole coupling for each atom in the structure (see SI).

| Method | NV-NMR | Bulk NMR | DFT |
|---|---|---|---|

| Material | $C_Q$ (kHz) | η | $C_Q$ (kHz) | η | $C_Q$ (kHz) | η |
|---|---|---|---|---|---|---|
| PMMA-d$_8$ | 57.12±1.15 | 0.024±0.015 | 51.8±1.05 | 0.052±0.016 | 187±3 | 0.056±0.027 |
| Phenanthrene-d$_{10}$ | 66.63±0.96 | 0.136±0.031 | 113.8±2.13 | 0.04±0.011 | 197±1 | 0.068±0.004 |

**Variable-temperature ²H NMR spectroscopy**

To probe molecular dynamics and phase change behavior further, we carried out variable-temperature NMR spectroscopy. Molecular motions are expected to increase at increasing temperature and with a change of phase, leading to characteristic changes in the resulting ²H NMR spectra (*38*).

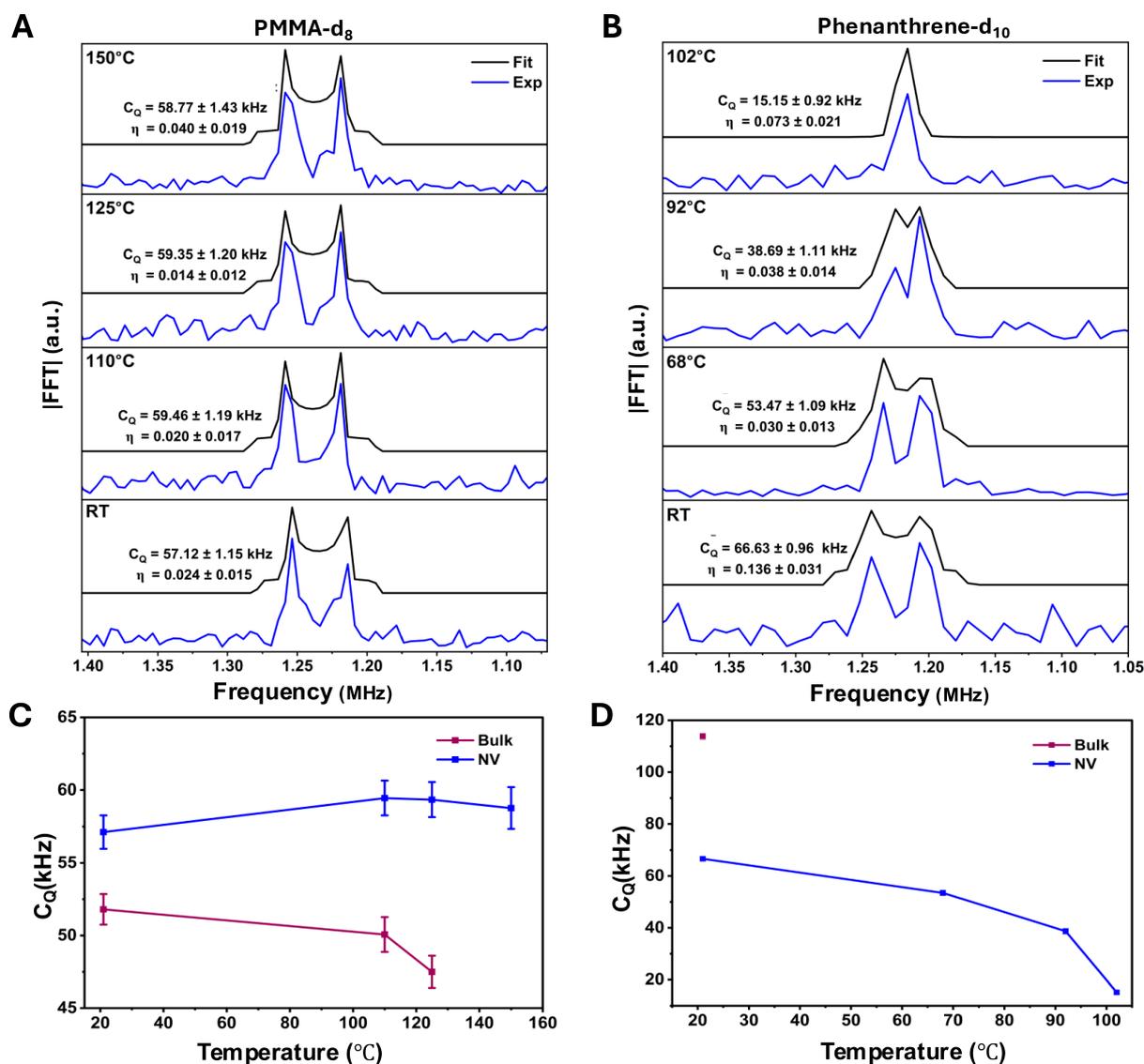

**Figure 3. Variable-temperature ²H NMR measurements. (A–B)** Variable-temperature NV-detected ²H NMR spectra (blue) of PMMA-d$_8$ (**A**), phenanthrene-d$_{10}$ (**B**), acquired at different temperatures (RT to elevated temperatures, as indicated). Experimental spectra are overlaid with fitted spectra (black), enabling extraction of $C_Q$ and η at each temperature. While PMMA-d$_8$ retains a largely temperature-independent quadrupolar splitting up to 150 °C, phenanthrene-d$_{10}$ exhibits progressive motional averaging with increasing temperature, reflected in reduced and eventually collapsed quadrupolar features upon melting. (**C**) Temperature dependence of the extracted $C_Q$ for PMMA-$d_8$ comparing NV-based (blue) with conventional bulk NMR (purple). For phenanthrene-$d_{10}$ **(D)**, NV-NMR shows a continuous decrease of the $C_Q$ with increasing temperature, while a weak bulk NMR signal precludes an extended temperature-dependent analysis.

Figure 3 shows the NV-detected $^2$H spectra and extracted quadrupolar coupling constants for the deuterated samples as a function of temperature, together with the corresponding bulk $^2$H NMR measurements shown in the lower panels. For PMMA-$d_8$ (Fig. 3A), the quadrupolar powder pattern observed at room temperature remains essentially unchanged upon heating up to ~150°C, indicating that no significant change in molecular motion (*i.e.*, no phase transition) occurs over this temperature range. The absence of significant line narrowing or a reduction in splitting indicates that motions in the polymer remain restricted, consistent with suppressed dynamics and an elevated local glass-transition temperature at the diamond surface (*48*). For bulk NMR experiments (Fig. 3C, Fig. S3, and Table S2), the variable-temperature data shows a minor narrowing up to 130°C, in line with an earlier report for bulk PMMA-$d_5$ (*43*). Measurements at higher temperatures were not possible due to experimental limitations. Together, these results confirm that molecular dynamics in PMMA remain largely constrained across both bulk and interfacial (nanoscale) environments over the measured temperature range.

In contrast, for phenanthrene-$d_{10}$, NV-NMR reveals a pronounced temperature dependence of the $C_Q$, reflecting enhanced motional narrowing at elevated temperatures. Phenanthrene-$d_{10}$ (Fig. 3B) exhibits a progressive coalescence of the quadrupolar lineshape with increasing temperature: starting from a broad pattern at room temperature, the spectrum starts narrowing around ~68 °C and continues to collapse through ~92 °C and ~102 °C, accompanied by a pronounced reduction in $C_Q$ (Fig. 3D). The melting point of phenanthrene-$d_{10}$ is reported to be approximately ~100°C according to supplier data, indicating that the observed spectral collapse occurs as the system approaches the solid–liquid transition, where molecular mobility increases significantly. In contrast, poor signal-to-noise and long averaging times for bulk phenanthrene $^2$H NMR measurements precluded a comparable temperature-dependent analysis with conventional NMR.

**DISCUSSION**

In this work we demonstrate nanoscale detection of $^2$H NMR signals, providing direct access to molecular motion and dynamics at the nanoscale. This represents an important step toward probing molecular dynamics, for example at interfaces or in confined samples. Recently, NV-based $^1$H NMR has been applied to probe confined water dynamics and interfacial phase behavior at solid surfaces, including the observation of liquid–solid transitions of nanoconfined water using quantum diamond probes (*49–52*). Detecting $^2$H instead of $^1$H can be a highly advantageous alternative, as the signal is free of background contributions from ubiquitous protons and provides insight into molecular dynamics via the quadrupolar interaction.

We emphasize that the intrinsic sensing capabilities of NV centers are a natural match for detecting quadrupolar nuclei. Consequently, low- or even zero-field detection of quadrupolar spins is likely to emerge as a key application domain of NV-based NMR (*53, 54*). Further applications include probing the molecular dynamics of lipids (*55*) or surface-tethered molecules (*56*), such as catalysts. Nanoscale detection of $^2$H dynamics could also find key applications in materials science, where questions of charge transport, dynamic motion at solid-state interfaces or surfaces, and reaction kinetics can be addressed. Additionally, in the case of strongly paramagnetic samples, NV-NMR offers low-field sensing capabilities, helping to mitigate the strong paramagnetic-induced broadening for solids at high fields while preserving sensitivity and spectral resolution. Finally, the sensitivity of our NV-NMR approach remains below its fundamental limit, leaving significant room for improvement through advanced NV readout techniques (*9, 57, 58*) and continued diamond engineering (*59, 60*).

In summary, we demonstrate nanoscale detection of $^2$H using NV-NMR spectroscopy, yielding quantitative quadrupolar line shapes that closely reproduce conventional bulk solid-state $^2$H NMR spectra – a long standing goal in the NV-NMR community (*21, 30, 61*). NV center–based detection offers two key advantages: operation at low magnetic fields and a sensitivity enhancement of six to eight orders of magnitude compared to conventional NMR, arising from the nanoscale detection volume. This capability paves the way for probing molecular dynamics and orientations at interfaces and, in the ultimate limit, in single molecules (*30*). Few, if any, other spectroscopic techniques can offer such insights.

## MATERIALS AND METHODS

**Sample preparation.** Prior to film deposition, the diamond substrate was cleaned with concentrated sulfuric acid to remove organic contaminants and ensure a clean, reproducible surface (*40*). After the acid treatment, the diamond was thoroughly rinsed and heated by a heat gun gently to eliminate residual moisture. The sample material was then directly placed onto the diamond surface, followed by gentle heating with a heat gun to ensure stable attachment of the sample to the diamond, resulting in a uniform film suitable for subsequent measurements.

**NV-diamond chip.** For the experiment, we utilized a custom-engineered diamond (Spinoway from Diatope GmbH) sensor. The chip consists of an electronic-grade, (100)-oriented, isotopically enriched $^{12}$C diamond substrate ($^{13}$C < 0.01%) with dimensions of $2 \times 2 \times 0.5$ mm³, hosting a near-surface ensemble of ~500–1000 NV centers per µm² at a depth of 5–10 nm. To enhance spin coherence and stabilize the near-surface environment, the diamond undergoes a proprietary surface treatment, resulting in reduced magnetic noise and improved charge stability.

**Surface NV-NMR setup.** A custom-built NV-NMR spectrometer, adapted from established designs (*27, 40*), was implemented with improved microwave control. The diamond chip was mounted on a thin glass coverslip (VWR 48393026 coverslip) and positioned between two neodymium magnets. By rotating and tilting the magnets, the static magnetic field $B_0$ (typically 170–190 mT) was aligned with one of the NV crystallographic orientations. NV centers were optically polarized and read out using a 532 nm laser (Verdi G7, Coherent) delivering ~250 mW to the diamond and focused to a ~40 µm diameter spot on the diamond surface. Laser pulse timing was defined by an acousto-optic modulator (Gooch & Housego, model 3260–220), yielding 5 µs pulses. To improve photon collection, a 6-mm BK7 glass half-ball lens (Edmund Optics, TECHSPEC) was attached to the bottom side of the coverslip. This configuration enabled coupling of the focused excitation laser (LA1986-A-M, Thorlabs) in a total internal reflection geometry. The resulting PL was collected and focused by two condenser lenses (ACL25416U-B, Thorlabs) and detected on an avalanche photodiode (Laser Components GmbH, ACUBE-S3000-10). The PL was filtered using a long-pass filter (Semrock, Edge Basic 647). The signal was digitized using a multifunction data acquisition card (National Instruments, USB-6229). An arbitrary waveform generator (AWG 5202,

Tektronix) was used to define the experimental timing and synthesize the microwave pulse sequences. A microwave source (SMB100A, Rohde & Schwarz) was mixed with the AWG output using an IQ mixer (MMIQ-0218LXPC, Marki Microwave) to generate the desired microwave frequency. The resulting signal was amplified by a high-power amplifier (ZHL-16W-72+, Mini-Circuits) and delivered to the diamond via a microwave loop antenna (*40*) positioned in close proximity to the sample for efficient spin control. The experiment is controlled by a custom-written MATLAB code (*11*).

**NV-NMR measurements.** The magnetic field at the NV center was determined by identifying the electron spin resonance (ESR) from the PL contrast. Experiments were performed in a magnetic field range of 174 mT-189 mT, which corresponds to $^2$H resonance frequencies between approximately 1.12 MHz-1.25 MHz. Fine control of the static magnetic field $B_0$ was achieved by adjusting the relative spacing between the magnets. Rabi experiments were used to calibrate the MW control pulses, yielding $\pi$ and $\pi/2$ pulse durations of ~36 ns and ~18 ns, respectively. Correlation spectroscopy measurements were performed using XY8-4 pulse blocks, corresponding to a total of 32 $\pi$ pulses, and sweeping the correlation time $t_{corr}$ starting from 3 µs to acquire the spectra. For PMMA-$d_8$, the $t_{corr}$ sweep was extended up to 203 µs, resulting in a total correlation time of 200 µs sampled with 1001 points. The acquired time-domain data were Fourier transformed, and the absolute value of the resulting spectrum was plotted using MATLAB. The $^2$H NMR signal of PMMA-$d_8$, shown in Fig. 2A, exhibits a signal-to-noise ratio (SNR) of ~11, calculated as the ratio of the signal amplitude to the root-mean-square (RMS) noise level evaluated in a signal-free region of the spectrum. The total experimental acquisition time for PMMA-$d_8$ was approximately 22 hours. For phenanthrene-$d_{10}$, a total correlation time of 110 µs sampled with 551 points was used, yielding a SNR of ~5, as shown in Fig. 2C, with a total experimental time of approximately 20 hours.

**Sensitivity estimation.** Spin sensitivity is defined as the minimum number of spins required to achieve SNR = 3 in 1 s integration time, $N_{min} = 3 \times \frac{N_{meas}\sqrt{T}}{SNR}$ $spins/\sqrt{Hz}$ assuming white-noise scaling. For the NV-based measurement, an SNR of ~11 was obtained after T = 22 h (79,200 s) from an effective 16 nm thick PMMA-$d_8$ film (detected by the NV-centers) and an area which is defined by the laser spot (r = 20 µm). The geometrical detection volume is V = $\pi r^2 t$, with t=16 nm denoting the effective sensitive thickness, yielding V = 2.01×10$^{-11}$ cm³. Using the PMMA density $\rho$ = 1.18 g cm$^{-3}$, repeat-unit molar mass M = 108.17 g mol$^{-1}$, Avogadro's number $N_A$ = 6.022×10$^{23}$ mol$^{-1}$, and 8 deuterons per repeat unit, the effective detection volume contains $N_{meas}$ ($^2$H) = 1.06×10$^{12}$ spins. This corresponds to an NV spin sensitivity of $N_{min}$ (NV) ~ 8×10$^{13}$ $spins/\sqrt{Hz}$. For comparison, a 50 mg (0.050 g) PMMA-$d_8$ bulk sample measured with SNR ~ 370 in T = 265 s contains 2.23×10$^{21}$ deuterons, corresponding to a bulk inductive NMR sensitivity of $N_{min}$ (bulk) ~ 3×10$^{20}$ $spins/\sqrt{Hz}$. This yields an improvement in spin sensitivity of 6–7 orders of magnitude for NV-based detection. For phenanthrene-$d_{10}$, we obtain $N_{meas}(^2H) \approx 8.0 \times 10^{11}$. With an SNR of ~5 in 20 h, this yields an NV sensitivity of $N_{min}(NV) \approx 1 \times 10^{14}$ $spins/\sqrt{Hz}$ The corresponding bulk measurement (50 mg, SNR ≈ 14 in 37 min 25 s) gives $N_{min}$(bulk) ≈ 2 × 10$^{22}$ $spins/\sqrt{Hz}$, corresponding to an improvement of ~8 orders of magnitude.

**Variable-temperature measurements.** We determined the sample temperature from the ODMR spectrum by fitting Lorentzian line shapes to the $m_s = 0 \rightarrow 1$ and $m_s = 0 \rightarrow -1$ transitions NV and using their mean frequency as a temperature-sensitive marker. First, a room-temperature (RT) reference spectrum is recorded, and the average of the two fitted resonance centers defines the baseline frequency $f_{rt}$ at 27.5°C. For any subsequent measurement, the current average resonance $f_{means}$ is computed in the same way, and the temperature change is obtained from the frequency shift |$f_{means}-f_{rt}$| using the reported coefficient of −74 kHz/°C (*62*). This procedure allows rapid thermometry directly from the ODMR data, with the initial RT dataset serving as the reference for all subsequent measurements. We note that this linear relation is accurate near room temperature, but deviations occur at elevated conditions (*63*). We therefore conservatively assume an uncertainty of ±2 °C in this regime.

**Bulk $^2$H solid–state NMR spectroscopy.** Solid-state $^2$H NMR spectra were collected on a Bruker Avance NEO 500 NMR spectrometer ($B_0$ = 11.75 T, $\nu_L$ = 76.77 MHz) equipped with a 4 mm Bruker triple resonance (HXY) magic-angle spinning (MAS) probe. Samples were ground and packed into 4 mm zirconia rotors with zirconia caps. Non-spinning $^2$H NMR spectra were acquired at natural abundance using Hahn-echo ($\pi/2 - \tau - \pi$) and solid-echo experiments ($\pi/2_x - \tau - \pi/2_y$) with 4.6 μs π/2 pulse ($\nu_{rf}$ ≈ 54 kHz), recycle delays of 2 s (PMMA) or 0.1 s (phenanthrene), and 128 (PMMA) or 16k-64k (phenanthrene) co-added transients. The $^2$H NMR spectra were referenced to D$_2$O (δ($^2$H) ≈ 4.8 ppm). Temperature control was achieved using nitrogen flow from a BCU-X temperature control unit and maintaining a constant probe temperature. NMR data were processed using Topspin 4.3.0 and plotted in Origin 2024.

**$^2$H NMR fitting.** Fitting of NV and bulk NMR data was carried out using the Quadrupole fitting routine in ssNake V1.5 (*64*). Errors were estimated by systematically altering the final fit parameters to achieve a maximum reasonable tolerance for peak position, $C_Q$, and η. The ranges where suitable fits are achieved are then given as the error range. Fits were exported and plotted in Origin 2024.

**DFT calculations.** DFT calculations were performed using ORCA 6.1.0 (*65–67*). Molecules were first built and starting geometries were optimized in Avogadro 1.2 (*68*), before calculations were performed in ORCA. Prior to electric field gradient (EFG) calculations, input structures were subjected to a geometry optimization in ORCA using the hybrid generalized gradient approximation (GGA) PBE0 functional and def2-TZVP basis set with Becke-Johnson dispersion corrections applied (*69–75*). A frequency calculation was performed on the final output structure to ensure the geometry was a true minimum. EFG calculations were then performed on the optimized structures with PBE0 functional until convergence in EFG parameters was reached at the def2-QZVP basis set level. For PMMA, to balance computational cost and calculation accuracy, a series of polymeric structures of increasing chain length were built and subjected to this same optimization and calculation procedure until EFG parameters were reliably reproduced. Starting with the monomer, methyl methacrylate (MMA), and increasing chain sizes (MMA-1 to MMA-5), variability in output EFG parameters was minimized at a chain length of 5 units, and these values are reported herein.

**EPR spectroscopy.** EPR Spectroscopy was performed on a Magnettech MS-5000 benchtop EPR spectrometer (Freiberg Instruments) at the X-band (applied frequency 9.461 GHz) with 1 mW microwave power and 0.08 mT field modulation at 100 kHz. Measurements were performed under ambient conditions, with 2 scans, 120 s sweep time, centre field of 337.5 mT, and 4 mT sweep width. The EPR spectrum was fit using the EasySpin simulation package in MATLAB 2025b. (*76*).

**Acknowledgements**
We acknowledge support from the Bavarian NMR Center (BNMRZ) at TUM for bulk NMR measurements, thank Riddhiman Sarkar for his support and assistance. We also thank Dr. Kirti Singh for valuable suggestions on the project. **Funding:** D.B.B. acknowledges funding from the Deutsche Forschungsgemeinschaft (DFG, Grant No. 412351169) within the Emmy Noether Program, as well as support under Germany's Excellence Strategy (EXC 2089/1—390776260 and EXC-2111—390814868). D.S. acknowledges support from the DFG Walter Benjamin Program (SI 3263/1-1). **Competing interests:** D.B.B. is cofounder and advisor of QuantumDiamonds GmbH, a company developing and commercializing NV sensing technology. The company did not have any relationship with this study. The remaining authors declare no conflict of interests. **Author contributions:** D.B.B. conceived and supervised the study. D.S. designed and performed experiments and analyzed the experimental NV data. R.W.H. performed the bulk NMR measurements and simulations. C.F. provides custom engineered diamonds, U.B. provided support for the NV-based measurements. D.S., R.W.H., and D.B.B. analyzed and discussed the data and wrote the manuscript with inputs from all authors. **Data and materials availability:** All data needed to evaluate the conclusions in the paper are present in the paper and/or the Supplementary Materials.

# Supplementary Information for

# Quantum sensing-enabled deuterium NMR spectroscopy with nanoscale sensitivity at low magnetic fields

Dileep Singh[1,2], Riley W. Hooper[1,2], Christoph Findler[3], Utsab Banerjee[1,2], Dominik B. Bucher[1,2*]

[1]Technical University of Munich, TUM School of Natural Sciences, Chemistry Department, Lichtenbergstraße 4, 85748 Garching bei München, Germany

[2]Munich Center for Quantum Science and Technology (MCQST), Schellingstr. 4, 80799, München, Germany

[3]Diatope GmbH, DLR-Building 3, Wilhelm-Runge-Straße 10, 89081 Ulm, Germany

*Corresponding author: dominik.bucher@tum.de

# Table of Contents

# Bulk $^2$H NMR Spectroscopy

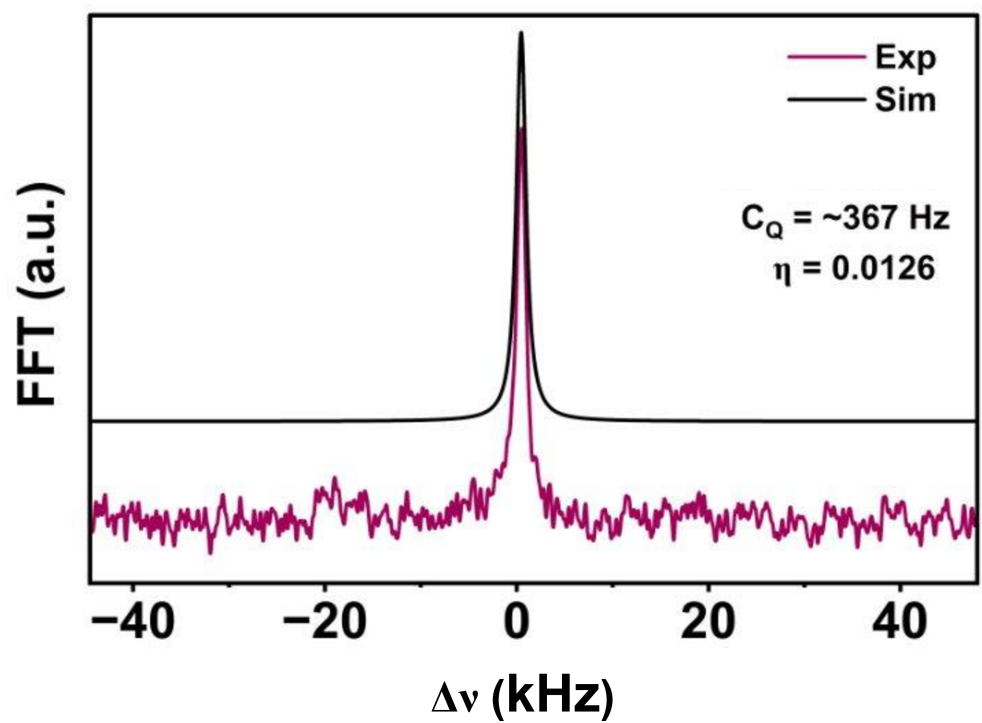

**Figure S1**. Hahn-echo detected $^2$H NMR spectrum for the narrow resonance of phenanthrene-$d_8$. $B_0$ = 11.75 T, $v_r$ = 0 kHz.

**Table S1**. Room temperature $^2$H quadrupolar fit parameters for phenanthrene bulk NMR data.

| Component | Δν (kHz) | $C_Q$ (kHz) | η |
|---|---|---|---|
| Broad (solid echo) | 24.54 | 113.8 ± 2.13 | 0.042 ± 0.011 |
| Narrow (Hahn echo) | 0.4693 | 0.3679 ± 0.98 | 0.013 ± 0.022 |

# EPR Spectroscopy

The spectrum exhibits a characteristic single resonance consistent with an organic radical in the phenanthrene-d$_{10}$ sample, with $g_{iso} \approx 2.0033$.

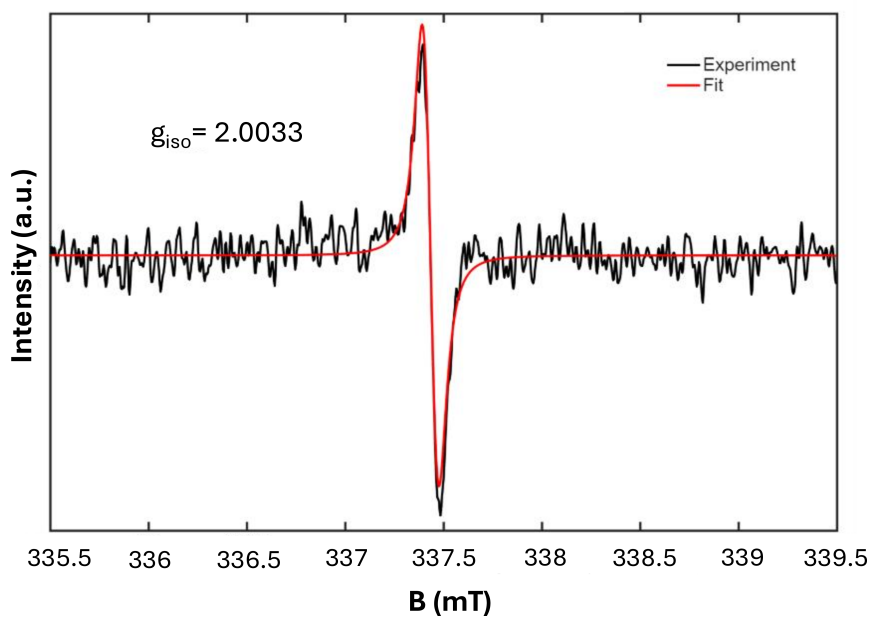

**Figure S2.** EPR spectrum at X-band (9.461 GHz) showing experimental data (black) and EasySpin fit (red).

# Variable-temperature $^2$H bulk NMR Spectra

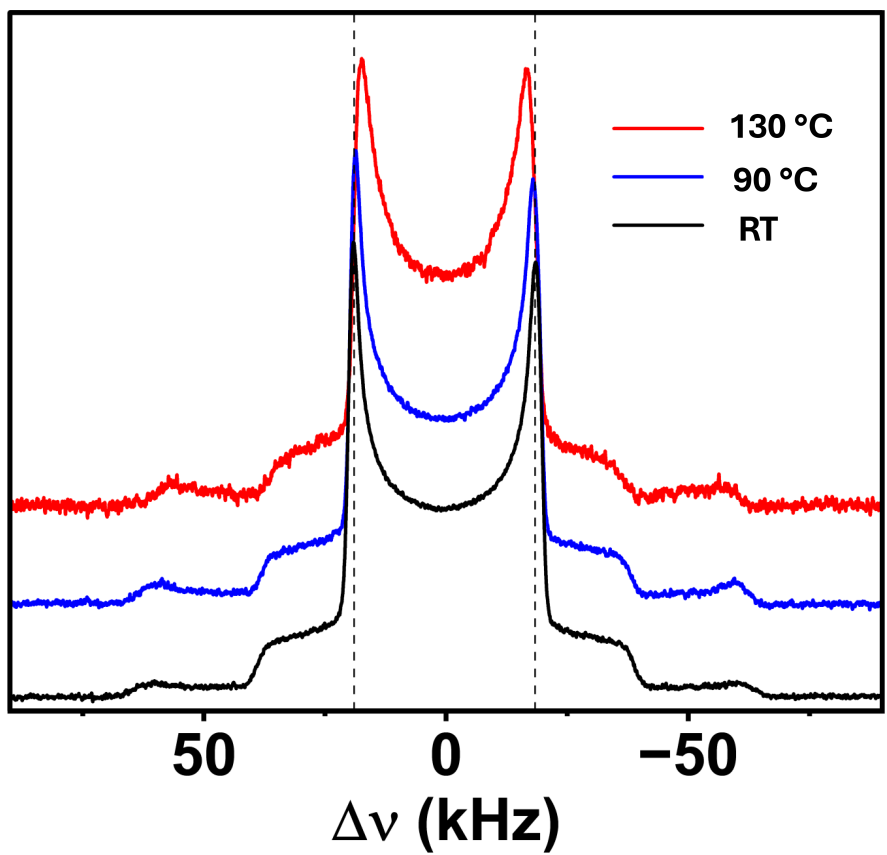

**Figure S3**. Variable temperature $^2$H NMR spectra for PMMA-d$_8$. $B_0$ = 11.75 T, $v_r$ = 0 kHz. Grey dashed lines are guides for the eye.

Extracted values from fitting of quadrupolar $^2$H variable-temperature NMR data for the CD$_3$ resonance are summarized below in Table S2.

**Table S2**. CD$_3$ quadrupolar fit parameters from variable-temperature PMMA-d$_8$ bulk $^2$H NMR data.

| Temperature (K) | Δν (kHz) | $C_Q$ (kHz) | η |
|---|---|---|---|
| 294 | 0.1640 | 51.8 ± 1.05 | 0.037 ± 0.016 |
| 363 | 0.1747 | 50.7 ± 1.20 | 0.042 ± 0.019 |
| 403 | 0.0943 | 47.5 ± 1.11 | 0.064 ± 0.025 |

# DFT Calculations

ORCA output geometry-optimized structures and calculated $^2$H quadrupolar parameters for PMMA and phenanthrene structures are summarized below. Detailed computational methods can be found in the MARERIALS AND METHODS section of the main text.

## PMMA

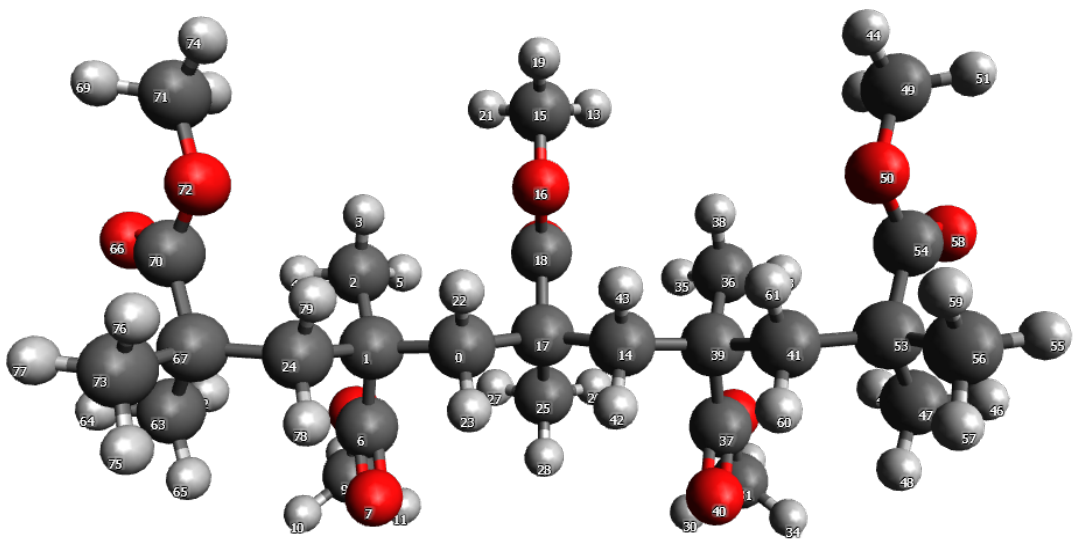

**Figure S4.** Geometry-optimized PMMA structure (MMA-5) consisting of five methyl-methacrylate units. Geometry optimized in ORCA 6.1.0 with PBE0 hybrid functional and triple-zeta basis set. Carbon = grey, hydrogen = white, oxygen = red.

**Table S3**. ORCA calculated $^2$H quadrupolar parameters for MMA-5. Atom labels correspond to hydrogen positions in the geometry-optimized structure. Parameters calculated with PBE0 hybrid functional and quadruple-zeta basis set.

| Atom | $C_Q$ (kHz) | η |
|---|---|---|
| 3H | 189.277 | 0.025499 |
| 4H | 188.071 | 0.028873 |
| 5H | 187.598 | 0.030033 |
| 10H | 184.752 | 0.095802 |
| 11H | 184.737 | 0.096074 |
| 12H | 189.019 | 0.083173 |
| 13H | 184.56 | 0.094981 |
| 19H | 189.154 | 0.082292 |
| 21H | 184.489 | 0.094775 |
| 22H | 183.854 | 0.059723 |
| 23H | 181.366 | 0.058414 |
| 26H | 189.301 | 0.035219 |
| 27H | 189.255 | 0.034886 |
| 28H | 187.476 | 0.031939 |
| 29H | 189.004 | 0.083296 |
| 30H | 184.654 | 0.09597 |
| 33H | 188.18 | 0.028606 |
| 34H | 184.765 | 0.095822 |
| 35H | 187.923 | 0.030989 |
| 38H | 189.322 | 0.024526 |
| 42H | 181.522 | 0.056751 |
| 43H | 183.733 | 0.060527 |
| 44H | 189.175 | 0.08259 |
| 45H | 189.224 | 0.034363 |
| 46H | 190.362 | 0.041185 |
| 48H | 188.087 | 0.028755 |
| 51H | 184.942 | 0.09508 |
| 52H | 184.447 | 0.095067 |
| 55H | 188.006 | 0.030785 |
| 57H | 188.659 | 0.030354 |
| 59H | 189.144 | 0.027478 |
| 60H | 181.435 | 0.051394 |
| 61H | 184.366 | 0.059177 |
| 62H | 189.261 | 0.034329 |
| 64H | 190.317 | 0.041172 |
| 65H | 188.113 | 0.028882 |
| 68H | 184.468 | 0.095051 |
| 69H | 184.944 | 0.095116 |
| 74H | 189.156 | 0.082581 |
| 75H | 188.674 | 0.030391 |
| 76H | 189.178 | 0.027356 |
| 77H | 188.05 | 0.030723 |
| 78H | 181.651 | 0.051833 |
| 79H | 184.074 | 0.059102 |
| **Avg** | 186.7669318 | 0.056384864 |
| **Std Dev** | 2.646487298 | 0.02714715 |

# Phenanthrene

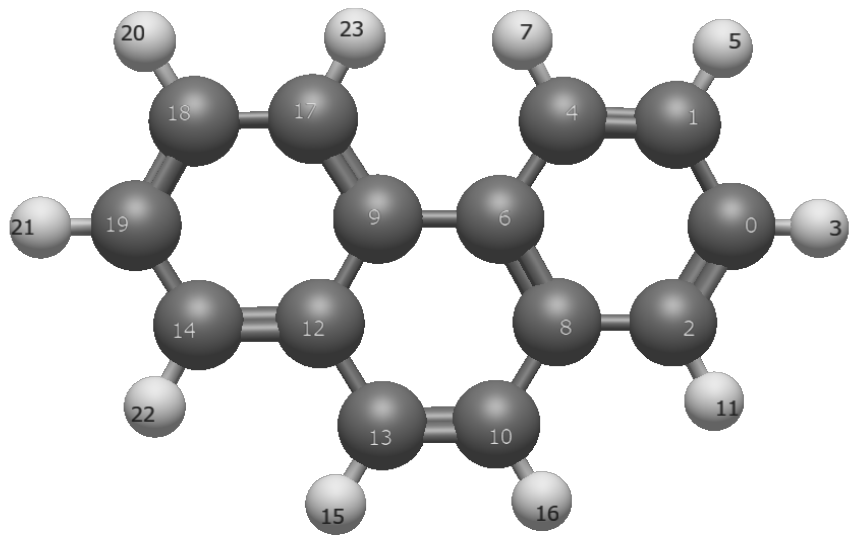

**Figure S5.** Geometry-optimized phenanthrene structure. Geometry optimized in ORCA 6.1.0 with PBE0 hybrid functional and triple-zeta basis set. Carbon = grey, hydrogen = white, oxygen = red.

**Table S4**. ORCA calculated $^2$H quadrupolar parameters for phenanthrene. Atom labels correspond to hydrogen positions in the geometry-optimized structure. Parameters calculated with PBE0 hybrid functional and quadruple-zeta basis set.

| Atom | $C_Q$ (kHz) | η |
|---|---|---|
| 3H | 196.856 | 0.062015 |
| 5H | 196.659 | 0.064415 |
| 7H | 198.57 | 0.073843 |
| 11H | 195.405 | 0.070171 |
| 15H | 195.493 | 0.069239 |
| 16H | 195.52 | 0.069228 |
| 20H | 196.678 | 0.064374 |
| 21H | 196.849 | 0.06204 |
| 22H | 195.404 | 0.070161 |
| 23H | 198.638 | 0.073768 |
| **Avg** | 196.6072 | 0.067925 |
| **Std Dev** | 1.1585 | 0.004209 |